 \documentclass[authoryear,preprint,review,12pt]{elsarticle}

\usepackage{amssymb}

\journal{Planetary Space Science}
\usepackage{url}
\begin{document}

\begin{frontmatter}



\title{On the estimation of the current density in space plasmas: multi versus single-point techniques}


\author[1]{Silvia Perri}
\author[1]{Francesco Valentini}
\author[2]{Luca Sorriso-Valvo}
\author[3]{Antonio Reda}
\author[1]{Francesco Malara}

\address[1]{Dipartimento di Fisica, Universit\`a della Calabria, Via P. Bucci CUBO 31C, I-87036 Rende, Italy}
\address[2]{Nanotec-CNR, Via P. Bucci CUBO 31C, I-87036 Rende, Italy}
\address[3]{Independent Scholar, Via Basso La Motta 2, I-87040 Mendicino, Italy}

\begin{abstract}
Thanks to multi-spacecraft mission, it has recently been possible to directly estimate the current density in space plasmas, by using magnetic field time series from four satellites flying in a quasi perfect tetrahedron configuration. The technique developed, commonly called ``curlometer'' permits a good estimation of the current density when the magnetic field time series vary linearly {\bf in space}. This approximation is generally valid for small spacecraft separation. The recent space missions Cluster and Magnetospheric Multiscale (MMS) have provided high resolution measurements with inter-spacecraft separation up to $100$ km and $10$ km, respectively.
The former scale corresponds to the proton gyroradius/ion skin depth in ``typical'' solar wind conditions, while the latter to sub-proton scale. However, some works have highlighted an underestimation of the current density via the curlometer technique with respect to the current computed directly from the velocity distribution functions, measured at sub-proton scales resolution with MMS.
In this paper we explore the limit of the curlometer technique studying synthetic data sets associated to a cluster of four artificial satellites allowed to fly in a static turbulent field, spanning a wide range of relative separation. This study tries to address the relative importance of measuring plasma {\bf moments} at very high resolution from a single spacecraft with respect to the multi-spacecraft missions in the current density evaluation.
\end{abstract}


\end{frontmatter}


\section{Introduction} \label{intro}

High resolution magnetic and plasma data in the interplanetary space have opened an important debate on the physical processes occurring between proton and electron scales. Indeed, spacecraft observations have revealed a steepening of the magnetic field power spectral density at scales at which the magnetohydrodynamics approximations are no longer valid, suggesting the presence of a small-scale turbulent cascade of magnetic energy \citep{Leamon98,Bale05,Sahraoui09,Alexandrova08,Alexandrova09,Sahraoui10}. This change of regime has been observed to occur at the proton gyroradius $\rho_p = v_{th,p}/\Omega_p$ ($v_{th,p}$ is the proton thermal speed and $\Omega_p$ the proton gyrofrequency) or at the proton skin depth $\lambda_p = c/\omega_p$ ($c$ is the speed of light and $\omega_p$ the proton plasma frequency). In addition to the spectral properties, it has been found that the plasma is characterized by magnetic discontinuities at proton scales and sub-proton scales both in the pristine solar wind \citep{Perri12,Greco16} and in the near Earth environment \citep{Retino07,Sundkvist07}. These evidences have raised the question about the role played by magnetic field discontinuities and current-sheet like structures in the magnetic energy dissipation. The general picture emerging from the analysis of high frequency spacecraft data is the coexistence of oblique propagating Kinetic Alfv\'en Waves and zero-frequency coherent structures, namely current sheets-like structures \citep{Roberts15,Perschke16}. 
On the other hand, three dimensional numerical simulations have pointed out the emergence of current sheets over a broad range of scales (up to electron scales) as a consequence of the development of magnetic turbulence; these are sites of high concentration of current density, energy dissipation, and plasma heating \citep{Karimabadi13, Wan15}. One of the possible processes responsible for this local energy dissipation is magnetic reconnection, a change in the topology of the magnetic field leading to a conversion of magnetic energy into heat, particle acceleration, and non-thermal effects \citep{Servidio12,Valentini14}.
Using high cadence measurements from the Magnetospheric Multiscale (MMS) mission, it has recently been possible to study in details electron scale magnetic reconnection and detect evidence of magnetic energy conversion into particle energy, electron currents, energy dissipation, and electron flows \citep{BurchEA16, Ergun16,Yordanova16,Fu17}.
Owing to sub-proton inter-spacecraft separations (i.e., minimum average distance $\sim 10$ km), MMS offers the best condition for the estimation of the current density applying the {\it curlometer} method \citep{Dunlop88,Dunlop02}. {\bf This technique was already applied to Cluster data during periods of inter-spacecraft separation of $\sim 200$ km \citep{Fu12}.}
Additionally, MMS measures three-dimensional plasma distributions with unprecedent time resolution, i.e., $150$ ms for ions and $30$ ms for electrons. Thus, high resolution current density can be derived from plasma moments as ${\bf J}_{mom}=qn({\bf V}_i- {\bf V}_e)$, where $q$ is the electric charge, $n=n_e=n_i$ is the plasma density, and ${\bf V}_i$, ${\bf V}_e$ are the ion and electron bulk speed, respectively \citep{Graham16}. The consistency between the current computed from multi-spacecraft and derived from the moments is generally satisfactory except for regions where structures at scales below the spacecraft separation are present \citep{BurchEA16,Graham16}.

In this paper, in order to explore the limit of validity of the multi-spacecraft approach, we apply the curlometer technique to a synthetic model of stationary turbulence \citep{Malara16,Pucci16} where four virtual spacecraft are allowed to fly forming a perfect tetrahedron with adjustable inter-spacecraft separation. We discuss the implications for possible future single-spacecraft missions in the estimation of the current density, which is of pivotal relevance in the study of plasma turbulence and dissipation.

\section{The curlometer technique}\label{curl}
The curlometer technique has widely been used in recent years thanks to multi-spacecraft missions. It is based on the the Maxwell-Amp\`ere's law $\mu_0 \bf{J}=\nabla\times \bf{B}$ evaluated in the centre of a perfect tetrahedron formed by four satellites (see Fig. 1 in \citet{Dunlop88}). Since this method is well known, here just a brief overview is given.
Starting from the ideal aforementioned configuration, one can estimate the current density in the direction normal to each face of the tetrahedron. Under the assumption that the magnetic field does not change abruptly over the inter-spacecraft separation, that is it varies linearly and the current density is roughly constant over the entire volume of the tetrahedron, one can write \citep{Grimald12}
\begin{equation}
\mu_0{\bf J}_{ijk} \cdot (\Delta {\bf r}_{ik}\times \Delta {\bf r}_{jk})=\Delta {\bf B}_{ik}\cdot \Delta {\bf r}_{jk}-\Delta {\bf B}_{jk}\cdot \Delta {\bf r}_{ik},
\label{jcurl}
\end{equation}
where $i,j,k$ are index running over the satellites, so that ${\bf J}_{ijk}$ is the current density normal to the face delimited by spacecraft $i,j,k$. $\Delta {\bf B}_{ik}={\bf B}_i-{\bf B}_k$ and $\Delta {\bf r}_{ik}={\bf r}_i-{\bf r}_k$ are the magnetic field and the position difference between spacecraft $i$ and $k$, respectively. Both the magnetic field data and the spacecraft positions are in Cartesian coordinates and an average current density in the tetrahedron, ${\bf J}_{\rm curl}$, can be derived by projecting each current vector normal to three faces into Cartesian coordinates. Because of the assumption of slow variation of the magnetic field, it is clear that the curlometer can be applied only when spacecraft are close enough to avoid sudden variation in the field inside the tetrahedron volume. This tends to limit the goodness of the estimation of the current density via this method.
To estimate the accuracy of the technique, one can evaluate $\nabla \cdot {\bf B}$, so that non-zero values are due to non-linear gradients in the magnetic field in the tetrahedron. Following, \cite{Dunlop88,Grimald12} we compute
\begin{equation}
div({\bf B})|\Delta {\bf r}_{ik}\cdot (\Delta {\bf r}_{jk}\times \Delta {\bf r}_{jl})=|\sum_{cyclic}\Delta {\bf B}_{ik}\cdot (\Delta {\bf r}_{jk}\times \Delta {\bf r}_{jl})|,
\label{divB}
\end{equation}
and in particular we calculate the adimensional quality factor
\begin{equation}
Q=\frac{div({\bf B})}{\mu_0{\bf J}_{\rm curl}},
\label{qfact}
\end{equation} 
so that $Q\ll 1$ indicates very good estimation of the current density via the curlometer.

\section{Application to synthetic data sets}\label{method}
\subsection{Numerical setup}
In order to test the limit of validity of the curlometer, we make use of a recently developed synthetic model of three-dimensional, static turbulence that reproduces the main characteristics of space plasma turbulence~\citep{Malara16,Pucci16}. 
The model mimics the turbulent cascade of energy from larger to smaller scales, and is based on an algorithm that allows to reproduce large spectral width and tunable spectral and intermittency properties, with very small computational requirements. A detailed description of the model can be found in~\citet{Malara16}, and a demonstration of its use can be found in~\citep{Pucci16}.
The main concept of the model is to build the magnetic field $\bf{B}(\bf{r})$ at each point $\bf{r}$ of the domain as the superposition of scale-dependent magnetic field fluctuations, chosen as to reproduce the desired turbulence characteristics. This is done by introducing a hierarchy of cells at different spatial scales $\ell_m=\ell_0/{2^m}$. Here $\ell_0$ is the domain size, $m=0,...,N_s$ is the scale index and $N_s$ is the tunable number of scales used in the realization, defining the spectral extension. At the largest scale, there is a single cell of size $\ell_0$. Then, each cell of a given scale is recursively divided into eight cells of the next scale size. At any given scale, the cells form a regular lattice filling the whole domain. Each cell in the model is indicated by four indexes $(i,j,k;m)$, where $i,j,k$ identify the cell position within the 3D lattice at the $m$-th scale. For each cell, a spatially localized magnetic field fluctuation ${\bf \delta B}^{(i,j,k;m)}({\bf r},\ell_m)$, or magnetic eddy, is univocally assigned through suitably defined polynomial functions and a series of random numbers that control the energy ``cascade''. Such random numbers determine the amplitude of each eddy, in order to reproduce both a given global energy spectral law and the desired amount of fluctuations inhomogeneity (or intermittency). 
The amount of energy transferred from larger to smaller scales shall obey a given power-law spectrum $E(k)\sim k^{\alpha}$, which implies that, on average, the field fluctuations scale as $\delta B(\ell)\sim \ell^{(\alpha-1)/2}$. For example, a Kolmogorov spectrum with $\alpha=-5/3$ is obtained allowing a scaling $\delta B(\ell)\sim \ell^{1/3}$. 
Intermittency is modelled as in the standard $p$-model~\citep{meneveau}, where $p\in [0.5, 1]$ is the intermittency parameter. In this model, the energy flows from larger to smaller eddies with a rate proportional to $p$ for some randomly selected cells, and to $1-p$ for the remaining cells. For $p = 0.5$, the energy transfer rate is homogeneously redistributed in the cascade, resulting in the absence of intermittency. As $p$ increases, the inhomogeneity of the energy transfer increases, enhancing the level of intermittency. Both the spectral index $\alpha$ and the intermittency level $p$ can be tuned as desired in the model. 
The magnetic field is finally calculated as 
%
%
\begin{equation}
{\bf B}({\bf r}) = {\bf B}_0 + \sum_{m=1}^{N_s} \sum_{i,j,k=1}^{2^m} {\bf \delta B}^{(i,j,k;m)}({\bf r},\ell)  \, ,
\label{eqB}
\end{equation}
where the fluctuations are determined following the above rules.

\begin{figure}
\includegraphics[width=12cm]{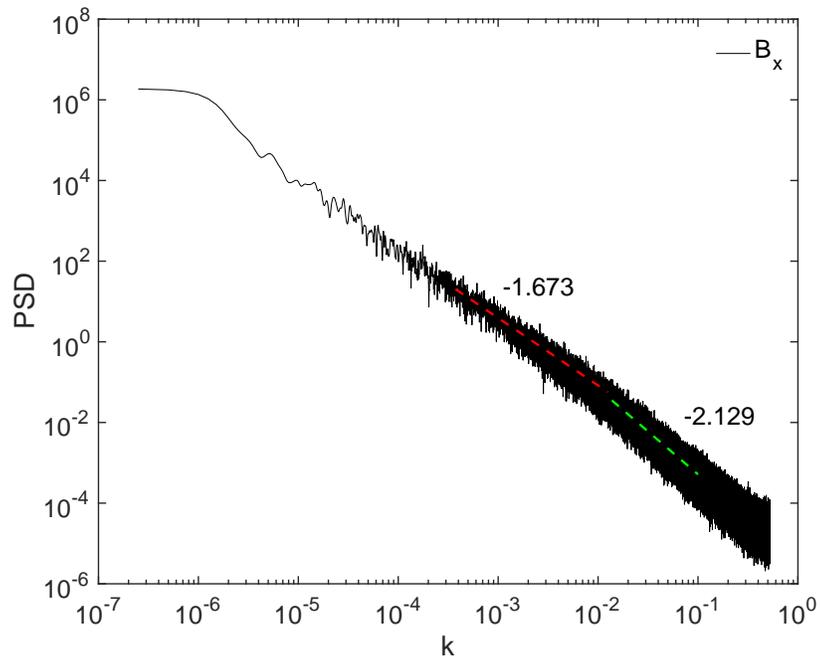}
\caption{Power spectral density of the $B_x$ component as a function of scales. The two power law ranges are highlighted by the dashed lines.}\label{figpsd}
\end{figure}

For the present work, the number of scales used in the model was $N_s=2^{22}$, allowing to have about six decades of spectral width. {\bf The largest scale chosen is equal to the typical correlation length at $1$ AU, namely $L\sim 5\times 10^6$ km \citep{Horbury96}, so that any spatial scale in the model is expressed in terms of $L$.}
The model was further modified as to allow the presence of a spectral break, i.e. two spectral indexes $\alpha_1$ and $\alpha_2$ are assigned above and below a given scale $\ell_{break}=122$ km. The two spectral indexes are chosen as to reproduce the double-scaling law observed in solar wind magnetic spectra, i.e.: $\alpha_1 = 5/3$ and $\alpha_2 = 2.1$~\citep{Alexandrova09}. 
The level of intermittency was set to $p=0.7$ in the large-scale range and to $p=0.5$ in the small-scale range.
An example of the spectrum of the $B_x$ component is shown in Figure \ref{figpsd}, where two power law ranges, one at large scales following a Kolmogorov-like scaling and one at smaller scales having $\alpha_2\sim -2$, are well resolved, as indicated by the thick dashed lines. 
Notice that since the form of the magnetic field ${\bf B}({\bf r})$ is analytically defined (see equation (\ref{eqB})), it is possible to exactly calculate the corresponding current density through the relation ${\bf J}=\nabla \times {\bf B}/\mu_0$.

\subsection{Estimation of the current density}
\begin{figure}[t]
\includegraphics[width=12cm]{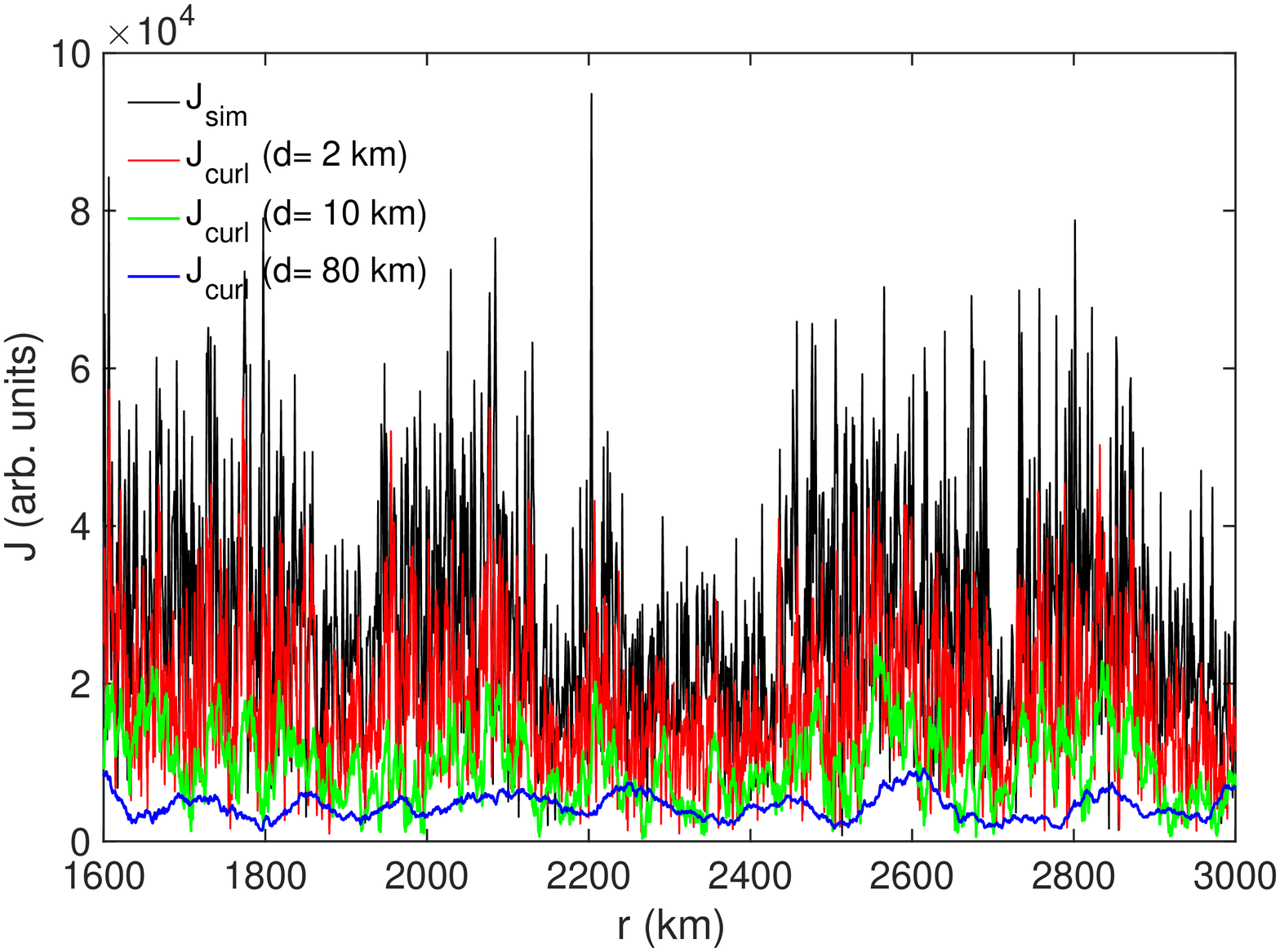}
\includegraphics[width=12cm]{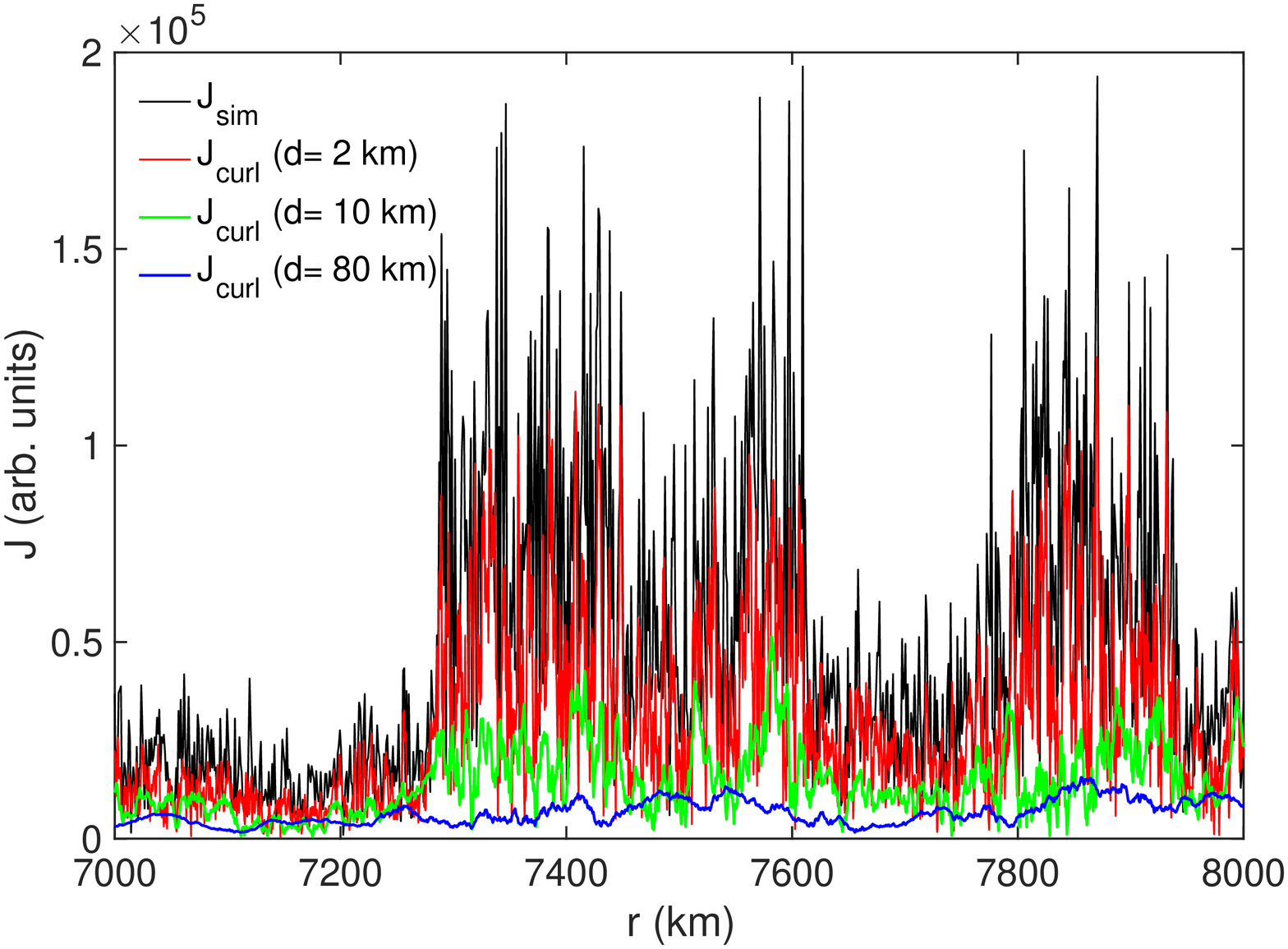}
\caption{Current density as a function of space computed from simulation (black line) and from the curlometer at several inter-spacecraft distances: $d=80$ km (blue line), $d=10$ km (green line), and $d=2$ km (red line), during a `quiet' period (top panel) and a `bursty' period (bottom panel).}\label{figcurl}
\end{figure}

\begin{figure}
\includegraphics[width=12cm]{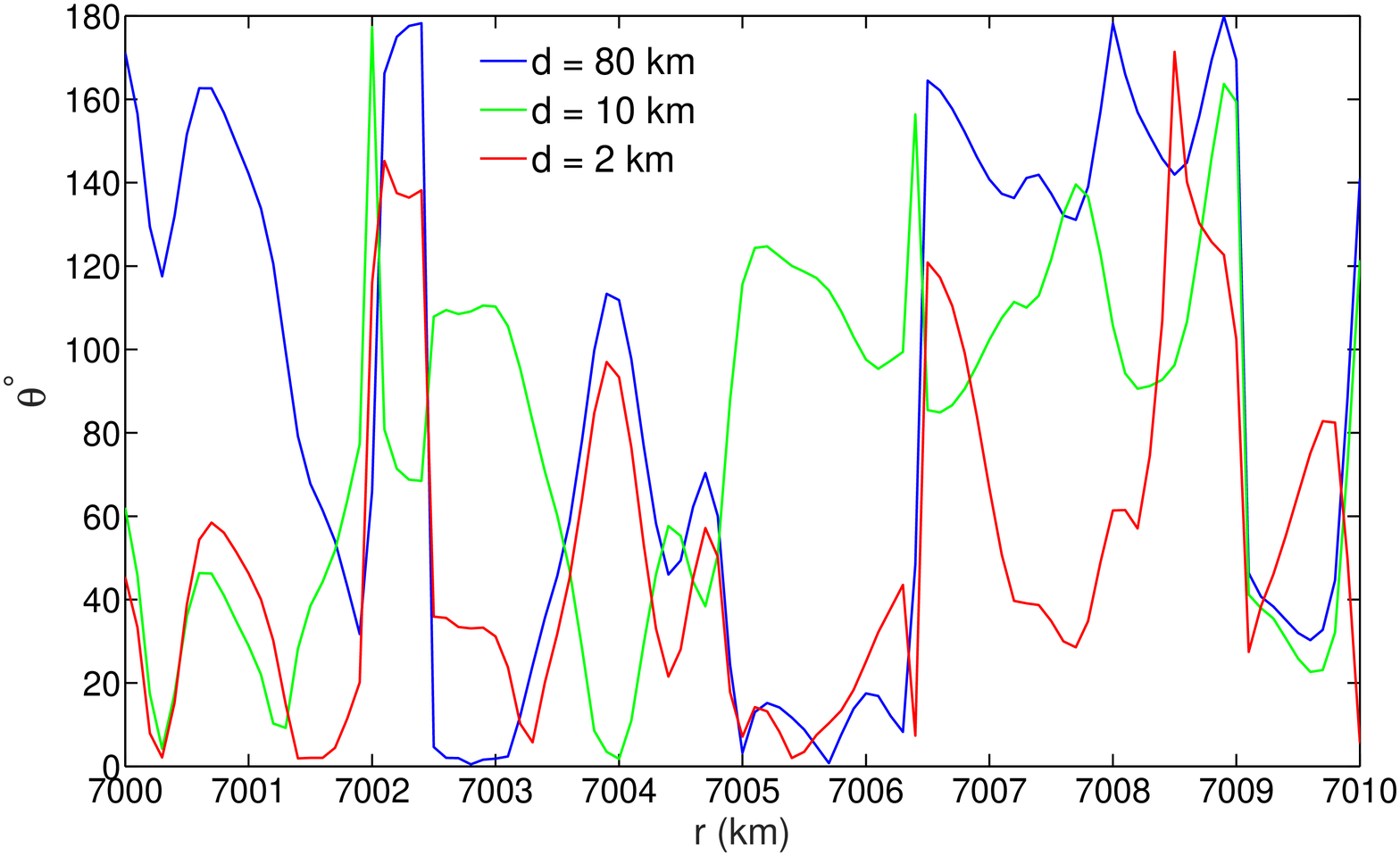}
\caption{The angle between the exact current computed via the Amp\'ere law and the current from the curlometer technique fixing the inter-spacecraft distance to $d=80$ km (blue line), $d=10$ km (green line), and $d=2$ km (red line). It measures the degree of alignment between the two currents. The greatest deviation from being aligned occurs for $d=80$ km (see text for more details).}\label{figangle}
\end{figure}

In this 3D turbulent field, it is possible to extract magnetic field time series as one-dimensional cuts of the $3$D field, simulating the flight of virtual spacecraft across the simulation domain. For our analysis, four virtual trajectories have been extracted at a certain relative separation, forming a perfect tetrahedron. With the appropriate choice of the model scale parameters, each virtual spacecraft measures the magnetic field at spatial cadence $\Delta s=0.1$ km. The current density is thus estimated using the spacecraft at different relative distance of $d=80$ km, $d=10$ km, and $d=2$ km. Notice that $80$ km corresponds roughly to the minimum distance reached by the Cluster spacecraft, while $10$ km to the minimum distance for the MMS mission. {\bf In order to compare the scales chosen for the estimation of $\mathbf{J}$ with physical scales, we arbitrarily consider to be immersed in a medium with average density of $n\sim 10 {\rm cm}^{-3}$, that is a typical value in the solar wind (see \citet{Alexandrova09}). In such a plasma, the electron inertial length is about $2$ km, close to the minimum distance chosen in our study for the artificial spacecraft. Notice that our model is purely magnetic and no plasma information can be derived.}
Figure \ref{figcurl} displays the current density estimated via the Amp\`ere's law $\mathbf{J}_{sim}=\nabla \times \mathbf{B}/\mu_0$ (black line), namely the exact estimation of $\mathbf{J}_{sim}$ in the numerical model, and via the curlometer (see Figure legend), i.e., equation (\ref{jcurl}). Top and bottom panels in Figure \ref{figcurl} correspond to a `quiet' period and to a more `bursty' one, respectively. The underestimation of the current density using the curlometer is evident at $d=80$ km and at $d=10$ km both in the bursty and in the quiet period. As expected, when strong gradient of the magnetic field are present, the curlometer technique fails because $\mathbf{B}$ does not change linearly across the tetrahedron. However, even using very short inter-spacecraft distances (i.e., $d=2$ km, which in typical solar wind conditions is close to the electron scales) the curlometer technique tends to underestimate the actual current density value. Besides a comparison between the {\bf magnitudes} of the current densities, we have also evaluated the degree of alignment between ${\bf J}_{sim}$ and the estimation of the current from the curlometer by computing
\begin{equation}
\theta({\bf r}) = \arccos\biggl[\frac{\mathbf{J}_{sim}({\bf r})\cdot \mathbf{J}_{curl}({\bf r})}{|\mathbf{J}_{sim}({\bf r})||\mathbf{J}_{curl}({\bf r})|}\biggr],
\label{eqangle}
\end{equation}
namely the angle between the current density vector calculated exactly and the current density vector from the curlometer. A close-up of $\theta({\bf r})$ is shown in Figure \ref{figangle}, where different colours refer to calculation of eq. (\ref{eqangle}) with $\mathbf{J}_{curl}({\bf r})$ estimated at different inter-spacecraft distances, i.e., $d=80$ km (blue line), $d=10$ km (green line), and $d=2$ km (red line). This quantity highly fluctuates, showing big deviations from $\theta=0^{\circ}$ in some regions, especially for the largest inter-spacecraft distance. Thus, we computed the spatial average value of $\theta$ in the region displayed in the top panel of Figure \ref{figcurl} (defined as the ``quiet'' period) obtaining $\theta_{(d=2 {\rm km})}=56^{\circ}$, $\theta_{(d=10 {\rm km})}=74^{\circ}$, $\theta_{(d=80 {\rm km})}=80^{\circ}$ (slightly higher values obtained for the ``bursty'' period). The higher the relative satellite distance, the bigger the deviation from alignment with the exact current density is found. These results highlight the limitations of multi-spacecraft technique in this kind of study.

\begin{figure}[h]
\includegraphics[width=12cm]{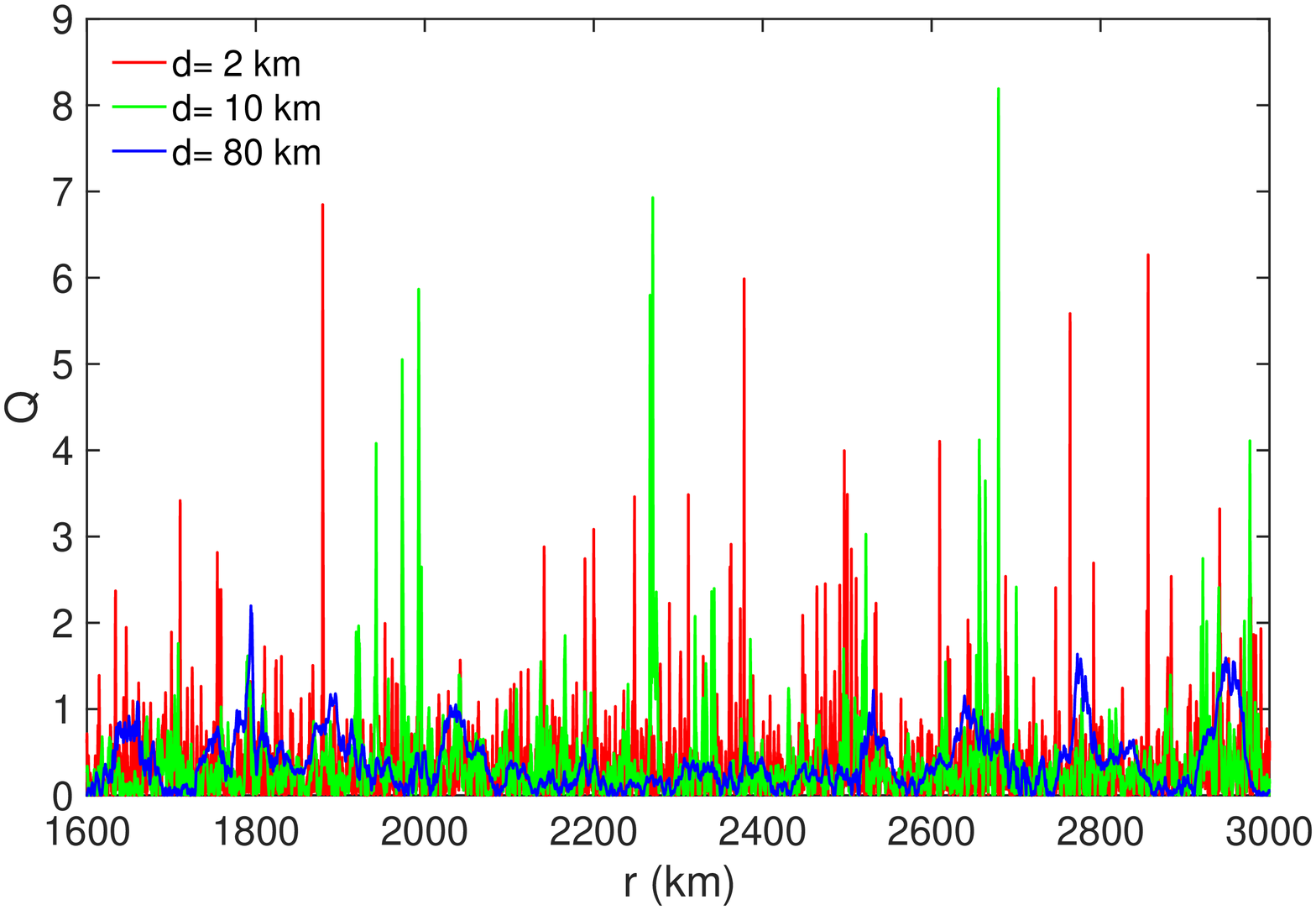}
\includegraphics[width=12cm]{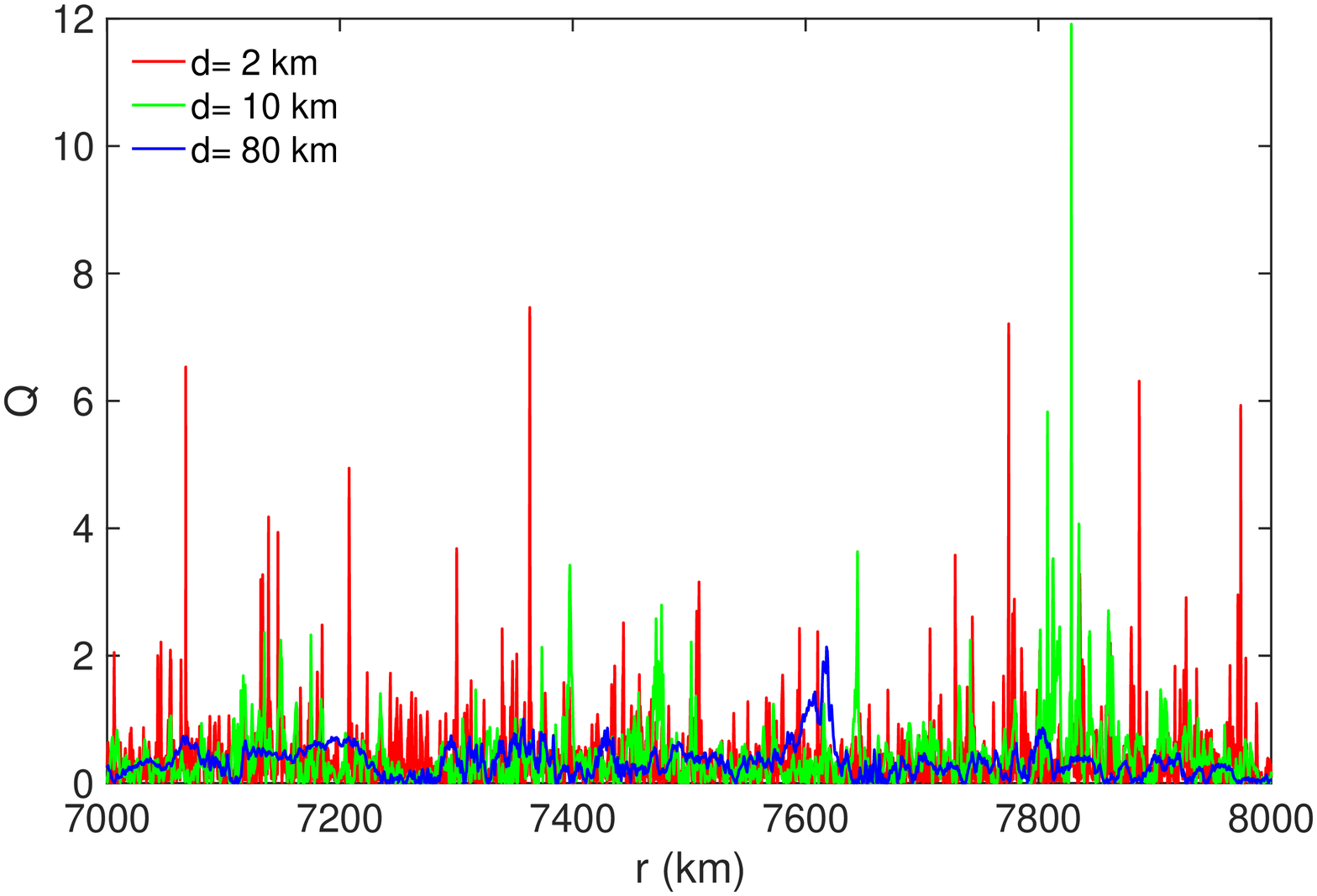}
\caption{Quality factor as a function of space computed for several inter-spacecraft distances: $d=80$ km (blue line), $d=10$ km (green line), and $d=2$ km (red line), during a `quiet' period (top panel) and a `bursty' period (bottom panel).}\label{figq}
\end{figure}

As stated in Section \ref{curl}, the quantitative evaluation on the goodness of the current density estimation has been done via the quality factor $Q$ (see equation \ref{qfact}), which is reported in Fig.\ref{figq} for the three satellite separations. It can be noted that frequently $Q\gg 1$ {\bf appear} both in bursty and quiet periods for all the three values of $d$. Large amplitude spikes are associated to points of almost vanishing current density (see Fig. \ref{figcurl}) and $\nabla \cdot {\bf B}>0$, {\bf in such regions $Q$ is not well determined.} Thus, up to $d=2$ km inter-spacecraft separation, the evaluation of the current density in several intervals can be poor. {\bf It is important to mention that recently, \citet{Fu15} investigated the limit of the quality factor computation in 3D kinetic simulations as a function of the separation among four virtual spacecraft. They also proposes an alternative method for testing the goodness of the reconstruction of a magnetic field topology (i.e., $\nabla \cdot \mathbf{B}$ evaluation).}
 
We have additionally estimated a relative error defined as $\langle{\rm Err_{J}}\rangle\equiv \langle|J_{curl}-J_{sim}|/J_{sim}\rangle$ during the intervals in Fig. \ref{figcurl}, where $\langle \;\rangle$ indicates spatial average. We obtained $\langle{\rm Err_{J}}\rangle_{d=2{\rm km}}= 0.52$, $\langle{\rm Err_{J}}\rangle_{d=10{\rm km}}= 0.69$, and $\langle{\rm Err_{J}}\rangle_{d=80{\rm km}}= 0.82$ in the bursty period (very similar value were also found for the almost quiet interval). Of course, the relative error increases on average with the inter-spacecraft separation.

\begin{figure}[t]
\includegraphics[width=12cm]{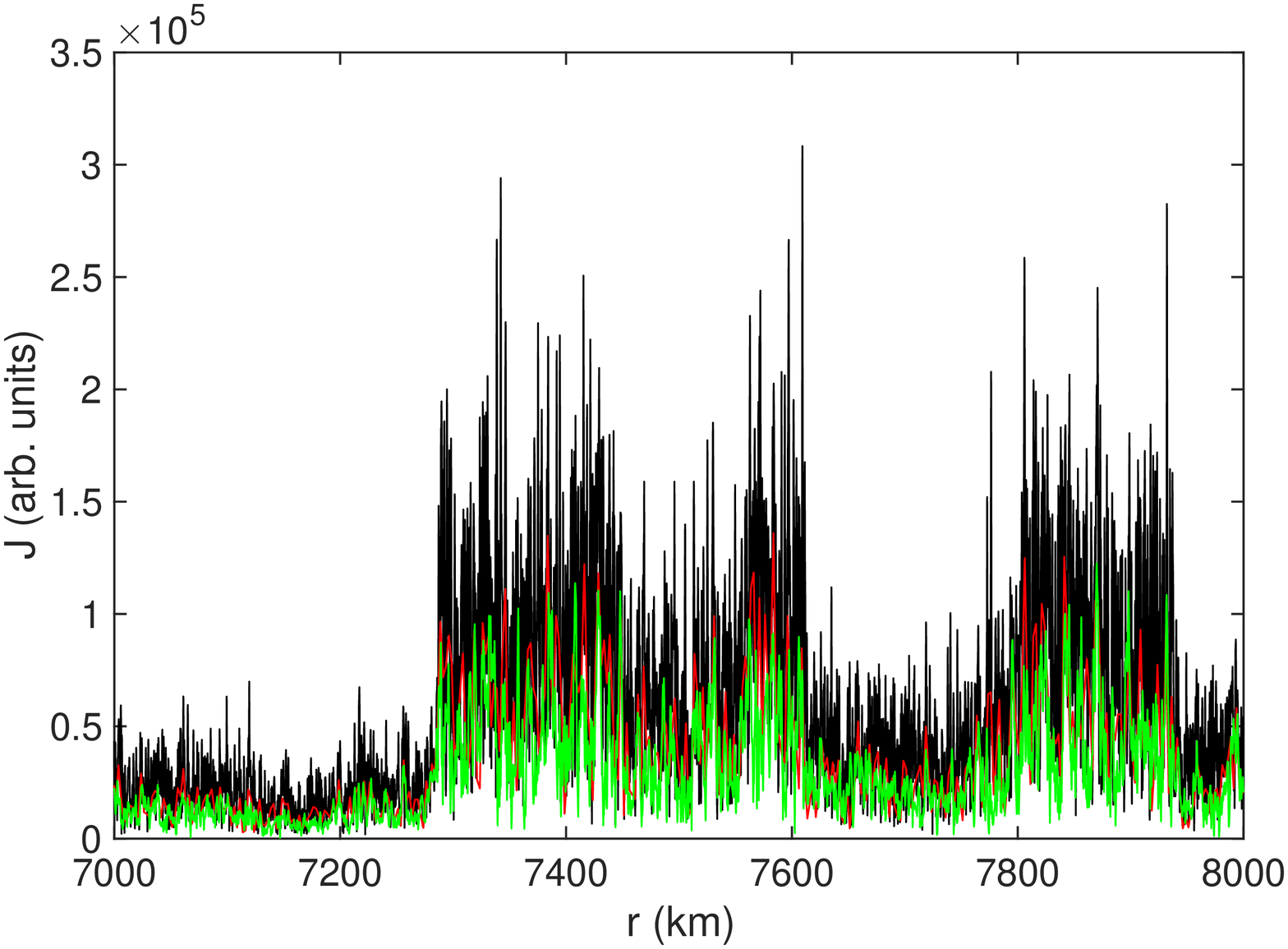}
\caption{Time series of the current density computed from a single artificial spacecraft along a trajectory in the 3D simulation domain at resolution $\Delta s=0.1$ km (black line) and $\Delta s=2.5$ km (red line). A comparison with the curlometer fixing $d=2$ km is reported (green line).}\label{figthor}
\end{figure}

Furthermore, we compare the exact current density estimation along a trajectory in the simulation domain at two spatial resolutions: $\Delta s=0.1$ km (black line in Fig. \ref{figthor}) and $\Delta s=2.5$ km (red line in Fig. \ref{figthor}). Assuming a typical solar wind speed of $V_{\rm sw}\sim 500$ km/s, the latter spatial lag corresponds to a time scale $\Delta t=\Delta s/V_{\rm sw}=5$ ms, which is the expected time resolution for the electron moments detection for the Turbulent Heating ObserveR (THOR) mission. As stated above, the exact estimation of the current density can be made via the measurements of plasma moments. Figure \ref{figthor} reports a comparison between {\bf $\mathbf{J}_{sim}$ along a path in the simulation box, that is} estimated with a single artificial spacecraft at high resolutions (see above) and via the curlometer at $d=2$ km (green line). It can be noted that high resolution single spacecraft measurements would lead to a better estimation of the current density than a multi-spacecraft mission, even with a unrealistic inter-spacecraft distance of $d=2$ km.

\section{Discussions and perspectives}\label{discussion}
The analysis performed on time/spatial series of magnetic field data from a synthetic turbulence model has pointed out that the evaluation of the current density via multi-spacecraft technique, as the curlometer, leads to frequent underestimations of this quantity, even in the presence of a relative small inter-spacecraft distance. This occurs because of the generation of {\bf sharp discontinuities} and structures towards small scales, as also recently highlighted from spacecraft data \cite{Perri12,Osman12,Greco16}. Regions of abrupt changes in the magnetic field over sub-proton scales can spoil the curlometer estimation. Thanks to high resolution MMS data, recent analysis has shown for the first time a direct comparison between the current density estimated from the curlometer technique (with inter-spacecraft separation of $10$ km) and the current density directly obtained from $150$ ms resolution plasma moments. This comparison has revealed a failure of the curlometer in the $\bf J$ estimation during some time intervals close to diffusion regions \citep{Graham16,Ergun16}. Our study on synthetic data {\bf clearly shows this limit of the multi-points technique, since} we would need four satellites with unrealistic (electron and sub-electron scales) minimum separation to get a good estimation of $\bf J$, owing to the presence of very narrow discontinuities in the magnetic field at small scales, possibly generated as the effect of the turbulent cascade. 
{\bf It is important to stress that the degree of deviation of the curlometer estimation from the exact $\mathbf{J}$ value depends on the occurrence of large amplitude gradients in the synthetic model. However, we do not want to compare directly our synthetic magnetic fluctuations with real magnetic data, but just to investigate and be predictive about the limit of this technique when a cluster of spacecraft crosses regions of sharp gradients in space plasmas even with a very short inter-satellite separation. It has been observed that in some space plasma regions the underestimation of the current density via the curlometer can be less dramatic (see Figure 3 in \citet{Graham16}). 

The results presented in this study indirectly suggest that the use of high-resolution single spacecraft plasma measurements would give a much better estimation of the current density (within the errors associated to the plasma moments determination) with respect to a multi-spacecraft mission. This should be particularly true in low beta plasmas, as the solar wind and the terrestrial magnetosheath.}  This point is of crucial importance to pursue THOR, one of the three candidates for the ESA's next medium-class science mission \citep{Vaivads16}.

\setcounter{secnumdepth}{0}
\section{Acknowledgements}
This work has been supported by the Agenzia Spaziale Italiana under the contract ASI-INAF 2015-039-R.O ``Missione M4 di ESA: Partecipazione Italiana alla fase di assessment della missione THOR

\end{document}